\documentclass[fleqn,10pt]{wlscirep}
\usepackage[utf8]{inputenc}
\usepackage[T1]{fontenc}
\usepackage{amsmath,amsfonts}
\usepackage{algorithmic}
\usepackage{algorithm}
\usepackage{array}
\usepackage[caption=false,font=normalsize,labelfont=sf,textfont=sf]{subfig}
\usepackage{textcomp}
\usepackage{stfloats}
\usepackage{url}
\usepackage{caption}
\usepackage{subcaption}
\usepackage{tikz}
\usepackage{subfig}
\usepackage{verbatim}
\usepackage{graphicx}
\usepackage{subfig}
\usepackage{multicol}
\usepackage{multirow}
\usepackage{makecell} 
\usepackage{booktabs,lipsum, longtable}
\usepackage{cite}
\usepackage{graphicx}
\title{Time-Domain Voice Identity Morphing (TD-VIM): A Signal-Level Approach to Morphing Attacks on Speaker Verification Systems}

\author[1,*]{Aravinda Reddy PN}
\author[2,+]{Raghavendra Ramachandra}
\author[1]{K.Sreenivasa Rao}
\author[1]{Pabitra Mitra}
\author[1]{Kunal Singh}
\affil[1]{Indian Institute of Technology Kharagpur,  India}
\affil[2]{Norwegian University of Science and Technology (NTNU), Norway}

\affil[*]{Corresponding Author: raghavendra.ramachandra@ntnu.no}

\keywords{\textit{Keywords--}Biometrics, Morphing attacks, Speaker verification systems, Voice biometrics, Smartphone biometrics}

\begin{abstract}
In biometric systems, it is a common practice to associate each sample or template with a specific individual. Nevertheless, recent studies have demonstrated the feasibility of generating "morphed" biometric samples capable of matching multiple identities. These morph attacks have been recognized as potential security risks for biometric systems. However, most research on morph attacks has focused on biometric modalities that operate within the image domain, such as the face, fingerprints, and iris. In this work, we introduce Time-domain Voice Identity Morphing (TD-VIM), a novel approach for voice-based biometric morphing. This method enables the blending of voice characteristics from two distinct identities at the signal level, creating morphed samples that present a high vulnerability for speaker verification systems. Leveraging the Multilingual Audio-Visual Smartphone database, our study created four distinct morphed signals based on morphing factors and evaluated their effectiveness using a comprehensive vulnerability analysis. To assess the security impact of TD-VIM, we benchmarked our approach using the Generalized Morphing Attack Potential (G-MAP) metric, measuring attack success across two deep-learning-based Speaker Verification Systems (SVS) and one commercial system, Verispeak. Our findings indicate that the morphed voice samples achieved a high attack success rate, with G-MAP values reaching 99.40\% on iPhone-11 and 99.74\% on Samsung S8 in text-dependent scenarios, at a false match rate of 0.1\%.

\end{abstract}
\begin{document}

\flushbottom
\maketitle
%
%
\thispagestyle{empty}
\noindent \textit{Keywords--}Biometrics, Morphing attacks, Speaker verification systems, Voice biometrics, Smartphone biometrics.

\section*{Introduction}

Voice biometrics is a crucial component of access control applications, where the user's voice is utilized for verification purposes. These systems are designed to identify unique features such as pitch, tone, cadence, and pronunciation from individual users to confirm their identity. Advancements in deep learning-based techniques have led to the emergence of voice biometrics as a reliable and accurate user verification tool. Unlike other biometric characteristics, voice biometrics are user-friendly, accessible, and highly accurate, making them suitable for use in banking and finance. Several prominent banking sectors, including HSBC \cite{bworld} and ING \cite{bworld}, have provided seamless banking services using voice biometrics for over a decade. In addition, these services have been extended to smartphones, leveraging the benefits of voice biometrics.

The implementation of voice biometrics in various applications has generated concerns regarding vulnerability to both direct and indirect attacks \cite{wu2015spoofing}. Direct attacks involve the presentation of voice artifacts from the voice capture sensor \cite{khan2023battling}, whereas indirect attacks involve the injection of voice artifacts into the system \cite{khan2023battling}. Direct attacks on voice biometric systems may take the form of presentation attacks, in which a voice sample corresponding to a legitimate user is presented to a voice biometric sensor using audio players \cite{wu2015spoofing}, such as smartphones, or through the use of deepfakes\cite{masood2023deepfakes} or voice synthesizers\cite{sanchez2015toward}, which are then played back to the sensor. Indirect attacks may involve the direct injection of voice artifacts into the voice system, with these artifacts potentially being either genuine or synthesized voice samples \cite{khan2023battling}.

Most existing attacks aim to compromise a legitimate user by generating a targeted artifact. However, adversarial attacks such as morphing, which seamlessly combines voice biometric characteristics from more than one identity, have raised concerns. The morphed voice sample can be used to match all identities whose voice samples are employed to generate morphing attacks, thus posing a high risk to application scenarios, such as banking and finance, where single identity verification is essential. Recently, Voice Identity Morphing (VIM) has been introduced as an attack vector that can be generated using voice samples from two different identities. The VIM method proposed in \cite{pani2023voice} is based on feature embedding and is extracted from two voice samples corresponding to two different identities. The features were extracted using a Deep Talk Encoder and were averaged to obtain the morphed feature embeddings. The averaged features are then used to synthesize the audio signal in two steps: (a) a Tacotron 2 speech synthesizer \cite{shen2018natural}  is used to generate the mel-spectrum from the averaged features with the help of the  reference text. (b) WaveRNN based neural vocoder \cite{kalchbrenner2018efficient}  is used to convert the melspectrogram into an audio sample that constitutes the VIM. The experimental results reported on two different speaker verification system (SVS) indicate the attack potential of feature-based VIM. However, the VIM proposed in \cite{pani2023voice}  has a few drawbacks: (a) it is highly dependent on the backbone used to extract the features, (b) feature to audio inversion requires reference text, and (c) it is limited to one type of language (i.e., English). Therefore, we are motivated to address these limitations to develop a more comprehensive and versatile VIM attack generation technique by proposing a novel technique to achieve a morphing operation in the time domain.

In this work, we propose a novel algorithm for VIM in which the morphing operation is performed in the time domain of voice samples. Given the voice signal uttering the same sentence from the two different identities, we first performed signal selection by considering different portions of signal corresponding to one of the identities while keeping another identity voice signal constant. We then perform averaging between the portions of the selected signal from one identity and the entire signal from another identity to render the TD-VIM signal. As morphing generation is performed in the time domain, the proposed approach is language and backbone independent and does not require reference text. We evaluated the attack potential of the proposed VIM using three different Speaker Verification System (SVS) on the publicly available dataset Multilingual Audio-Visual Smartphone (MAVS) dataset \cite{mandalapu2021multilingual}, which contains 103 data subjects captured using two different smartphones and three different languages. The main contributions of this study are as follows: 
\begin{itemize}
    \item We introduce Time Domain - Voice Identity Morphing (TD-VIM), a novel approach for generating morphed speech at the signal level. For this task we used publicly available databases Multilingual Audio-visual Smartphone dataset (MAVS). The TD-VIM approach enables seamless voice morphing directly in the time domain, allowing identity blending without any embeddings from the backbone, or reference text.  
    \item We evaluate the attack potential of TD-VIM on three Speaker Verification Systems (SVS): the x-vector, RawNet3, and a commercial off-the-shelf system, Verispeak. The analysis covers four types of morphing attacks for both text-dependent and text-independent cases (detailed results are available in the supplementary material).
\item Using the Generalized Morph Attack Potential (G-MAP) metric, we conduct extensive experiments to measure the attack success rate of TD-VIM across various devices and languages. This helps us determine if device type and language influence the vulnerability of morphed samples.

\item The morphed files and the original MAVS dataset are available from the corresponding author upon reasonable request. The SWAN database is publicly accessible and can be obtained via request at (\url{https://zenodo.org/records/3925170}) and to promote reproducibility and transparency, we provide the source code for TD-VIM in the repository (\url{{https://github.com/Aravinda27/TD-VIM}}).

\end{itemize}

The rest of the paper is organized as follows: Section 1 introduces our proposed method. Section 2 details the experimental protocol used to evaluate the method. Section 3 presents a vulnerability analysis of the proposed approach. Finally, Section 4 concludes the paper. 

\section{Proposed method: Time domain-Voice Identity Morphing (TD-VIM)}
\label{sec:proposed method}

\begin{figure*}[!ht]
\centering
\includegraphics[width=1\textwidth]{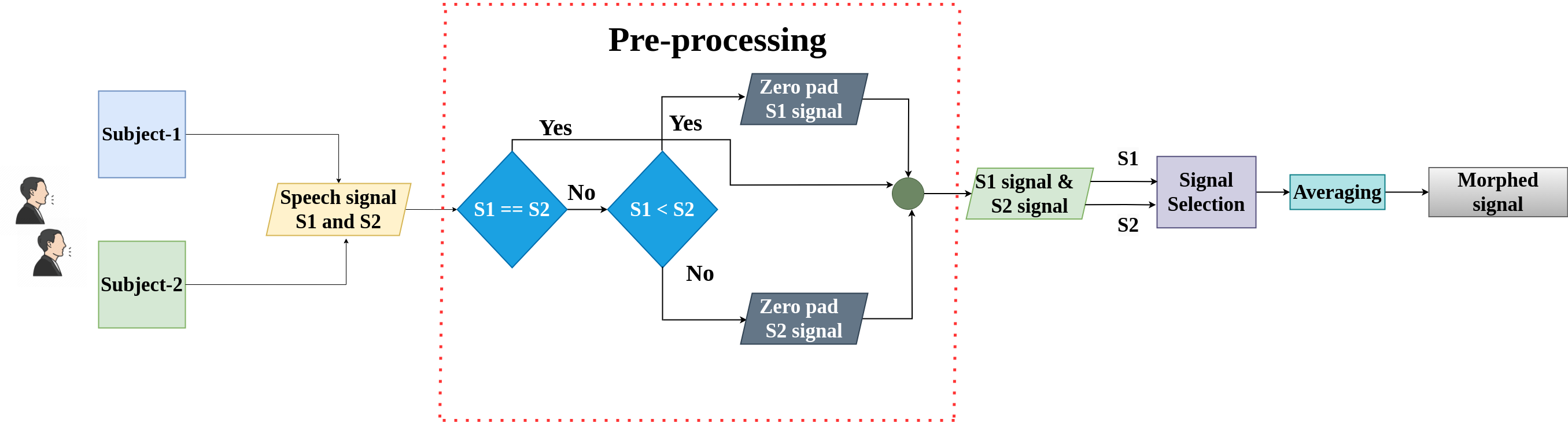}
\caption{Illustration of TD-VIM morphing: Initially we apply pre-processing to make
both signal of equal length. Subsequently we select four different portions of second signal in the signal selection block and
average only the selected portion with the first speaker's signal. This averaged signal is final morphed signal
which is used to verify both the subjects}
\label{fig:proposed_time}
\end{figure*}

\begin{figure*}[!h]
\centering
\includegraphics[width=\textwidth]{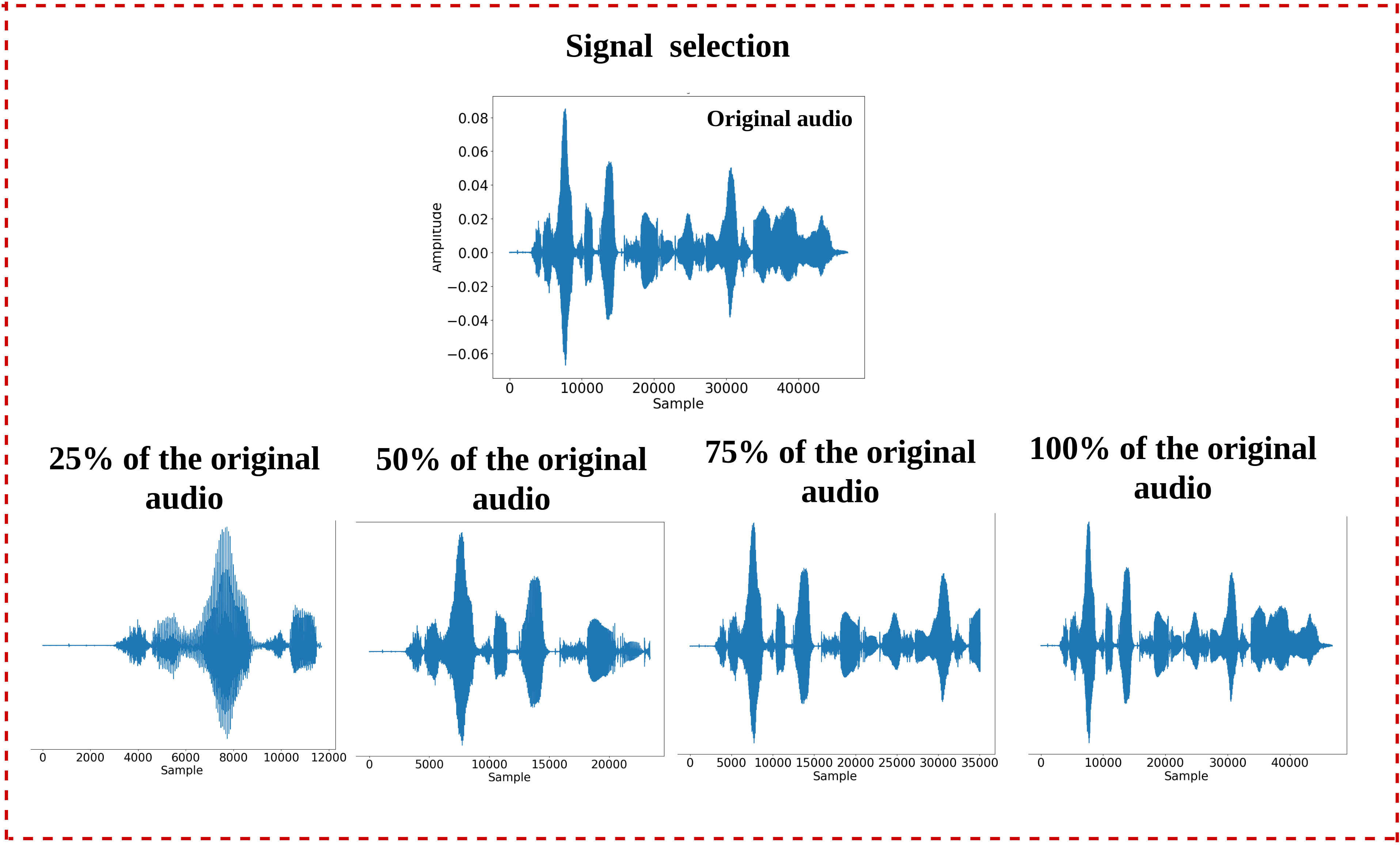}
\caption{Illustration of signal selection of second speaker, based on the time samples we have selected 25\%, 50\%, 75\% and 100\% of the second speaker}
\label{fig:signal_selection}
\end{figure*}

The proposed block diagram of the TD-VIM in the time domain is shown in thenjaNJDJ Figure \ref{fig:proposed_time}. The proposed morph generation process aims to select two contributory subjects based on gender and average signals in the time domain to generate a morphed speech signal. The proposed TD-VIM consists of five steps: \textbf{Speaker Selection:} Two contributory speakers are chosen from the database based on their gender and  both have uttered the same spoken content. \textbf{Pre-processing:} The lengths of the two speech signals are compared. The shorter signal was zero-padded to match the length of the longer signal. This ensured that both signals had the same duration and could be accurately processed together. Signal Selection: Since both speakers uttered the same content, we have selected four different portion of second  individual's speech (see Figure  \ref{fig:signal_selection}). \textbf{Averaging:} We average only the selected portion of speech signal from second contributory subject with the first contributory subject. This averaging process combines the acoustic characteristics of both speakers, resulting in a morphed speech signal that exhibits blended features of both individuals. \textbf{Verification: }Finally the obtained morphed signal is verified for both contributory subjects. In the following, we discuss each of these steps in details. 
    

\subsection{Speaker selection}
The proposed TD-VIM morph generation process is based on judicious selection of two contributory speakers from the database based on gender pair i,e., male-male (MM), female-female (FF) pairs having uttered the same spoken content. For this task, we have chosen MAVS database, which has three languages English, Hindi and Bengali \cite{mandalapu2021multilingual}.

\subsection{Pre-processing}
Given two speech signal of variable length we perform zero padding on the shorter duration signal. Let us denote subject-1 speech signal as $S_1(n)$ and subject speech signal as $S_2(m)$. Let us assume $S_1(n)$ with signal length N1 and $S_2(m)$ with signal length N2. Steps for zero padding includes:
\begin{itemize}
    \item Calculate the difference in length.
            \begin{equation}
            \centering
                D=|N1-N2|
            \end{equation}
    \item Determine the padding size:
    \begin{itemize}
        \item If $N1<N2$, pad the $S_1(n)$ with D zeros.
        \begin{equation}
        \centering
        N1 < N2 : 
        S_1(n) = \begin{cases}
            S_1(n) & \text{for $0 \leq n < N1$} \\
            0 & \text{for $N1 \leq n < N2$}
                \end{cases}
        \end{equation}
        \item If $N1>N2$, pad the $S_2(m)$ with D zeros.
        \begin{equation}
        \centering
            N1 > N2 : 
                S_2(m) = \begin{cases}
                    S_2(m) & \text{for $0 \leq m < N2$} \\
                    0 & \text{for $N2 \leq m < N2$}
                        \end{cases}
        \end{equation}

  
        

  
        
    \end{itemize}
\end{itemize}

\subsection{Signal selection}
Four portions of the second contributory signal were selected. Here, a breakdown of the different portions is based on the percentages (see Figure \ref{fig:signal_selection}):
\begin{itemize}
    \item 25\%: This portion represents the first quarter of the second contributory signal. 
    \item 50\%: This portion encompasses the first half of the second contributory signal. 
    \item 75\%: This portion covers the first three-quarters of the second contributory signal. 
    \item 100\%: This portion utilizes the entire second contributory signal.
\end{itemize}

\subsection{Averaging}

    Given two speech signals having same spoken content from two different speakers, we name the two speakers as subject-1 and subject-2 respectively. We denote subject-1 as $S_1$ full spoken content and other subject is named as $S_{2i}$ where $proportion={25\%,50\%,75\%,100\%}$ is the percentage of spoken content of second contributory speakers. We generate 4 different types of morphed signal is defined by:
\begin{enumerate}
    \item Select the proportion of the second signal $proportion$ to add to the first signal.
    \item Calculate the length of the selected portion of the second signal $p=int(N_2 \times proportion)$.
    \item Add the selected proportion of the second signal to the first signal:
    \begin{equation}
    \centering
        averaged\;signal[i]=\sum_{i=0}^{p-1}S_1[i]+S_2[i]
    \end{equation}
    \begin{equation}
    \centering
        S_{Morph}[i]=\sum_{i=0}^{N_1-1}\frac{averaged \;signal}{2}
        \label{eq:morph_gen}
    \end{equation}
\end{enumerate}

$S_{Morph}[i]$ is generated morphed speech signal for each value of i.
We denote $M_{25}$ for case-1, $M_{50}$ for case-2, $M_{75}$ for case-3, $M_{100}$ for case-4 morphing for our convenience. 

\section{Experimental protocol}
\label{sec:experimet}
\subsection{Datasets}
 We have selected publicly available database Multilingual Audio-visual Smartphone based dataset \cite{mandalapu2021multilingual} for morph generation process and additional results for SWAN multi-modal biometric dataset \cite{ramachandra2019smartphone} is also discussed in Supplementary material.

\subsubsection{Multilingual Audio-Visual Smartphone (MAVS) Dataset:}

The MAVS dataset comprises audio-visual biometric data collected from 103 subjects, including 70 males and 33 females, across various age groups. Table \ref{tab:my-table:MAVS} shows the statistics of the MAVS dataset. The data were gathered using five different smartphones, namely, iPhone 6s, iPhone 10, iPhone 11, Samsung S7, and Samsung S8, in three distinct sessions and three languages: English, Hindi, and Bengali. For the purpose of this study, we selected two devices, Samsung S8 and iPhone 11, and all speakers delivered six sentences each. Among these, we chose three sentences spoken by all the speakers. The following sentences are the common sentences used to create morph.

\begin{itemize}
    \item S1: I am working at IIT Kharagpur.
    \item S2: The limit of my account is 10,000 rupees.
    \item S3: The code for my bank is 9876543210.
\end{itemize}


\subsubsection{Baseline Verification Performance}
Before assessing the vulnerability of x-vectors, RawNet3 and Verispeak based SVS systems we evaluate the verification performance by computing the equal error rate (EER) which corresponds to False Matching Rate (FMR) is equal to False Non-matching Rate (FNMR) on MAVS dataset. For our experiment we have set an operating threhold of 0.5 for all the three SVS. The deep learning based SVS used in this work are x-vector \cite{snyder2018x} and RawNet with Additive Angular Margin (AAM) Softmax or popularly called ArcFace loss used in face recognition \cite{jung2022pushing} and we use a Commercial-off-the-shelf (CTOS) SVS such as Verispeak \cite{Verispeak}.
The x-vector based SVS uses Time Delayed Neural Networks (TDNN) to excerpt 512-dimensional deep embeddings by statistically poling the varying length sentences. The Rawnet3 ASV system based on raw waveform inputs has received a very little attention. The Rawnet3 based architecture is hybrid mixture of RawNet2 \cite{jung2020improved} and ECAPA-TDNN \cite{desplanques2020ecapa}. In RawNet3, the input signal is  pre-emphasised following this process, the boosted signal is passed to instance normalisation layer \cite{ulyanov2016instance}. Then the output of this layer is passed to the adaptive center frequency bandwidth based parameterised analytic filter banks \cite{pariente2020filterbank}. The parameterised filter banks are the extension of sinc convolutional layer \cite{ravanelli2018speaker} where sinc function is multiplied by a complex exponential and thereby the learned filters offer very fewer parameters and better interpretability. The output of parameterised filter banks are passed onto the three backbone blocks with residual connections called Advanced Feature Map Scaling with Res2MP blocks of Rawnet2 architecture \cite{jung2020improved} and at last max-pooling layer is applied to obtain the 256 dimensional embeddings. VeriSpeak SVS \cite{Verispeak} was designed for biometric system developers and integrators. The SVS algorithm ensures system security by checking both the voice and phrase authenticity. Voiceprint templates can be matched in 1-to-1 (verification) and 1-to-many (identification) modes. More details regarding VeriSpeak is shown in Supplementary material.

The verification performance of these systems are shown in Table 
\ref{tab:EER_MAVS}. The x-vector based SVS outperforms RawNet3 and Verispeak for all three languages. However for English language alone Verispeak performs better when compared to x-vector. Lower EER systems are generally considered as more robust in speaker verification, striking a better balance between authentic speaker acceptance and imposter rejection. The tables clearly demonstrate the chosen systems lower EERs, validating our selection.

\begin{table}[!h]
\centering
\footnotesize
\caption{Table showing exhaustive list of number of morphed files generated for 2 devices for 3 sentences for 1 sessions using 4 different types of morphing technique for 3 different languages }
\label{tab:my-table:MAVS}
\begin{tabular}{|ccc|c|}
\hline
\multicolumn{1}{|c|}{\textbf{Devices}} & \multicolumn{1}{c|}{\textbf{\makecell{No of \\ speakers}}}                  &     \textbf{\makecell{No of \\ sessions}}              & \textbf{\makecell{No of \\ morphed files}} \\ \hline
\multicolumn{1}{|c|}{iPhone 11} & \multicolumn{1}{c|}{\multirow{2}{*}{103}} & \multirow{2}{*}{1} & \makecell{$103$ $\times$ $102$ $\times$ $3$ $\times$ $4$ \\ $\times$ $3$ $=$ $378,216$}  \\ \cline{1-1} \cline{4-4} 
\multicolumn{1}{|c|}{Samsung S8} & \multicolumn{1}{c|}{}                  &                   & \makecell{$103$ $\times$ $102$ $\times$ $3$ $\times$ \\ $4$ $\times$ $3$ $=$ $378,216$}   \\ \hline
\multicolumn{3}{|c|}{}                                                              & Total=2,269,296 \\ \hline
\end{tabular}
\end{table}

\begin{table}[!h]
\centering
\caption{Performance of the speaker verification systems in terms of EER (\%) for MAVS 
database}
\footnotesize
\begin{tabular}{|cccc|}
\hline
\multicolumn{4}{|c|}{\textbf{MAVS Database}}                                                                     \\ \hline
\multicolumn{1}{|c|}{\multirow{2}{*}{\textbf{SVS}}} & \multicolumn{3}{c|}{\textbf{Languages}}                            \\ \cline{2-4} 
\multicolumn{1}{|c|}{}                  & \multicolumn{1}{c|}{\textbf{English}} & \multicolumn{1}{c|}{\textbf{Hindi}} & \textbf{Bengali} \\ \hline
\multicolumn{1}{|c|}{\textbf{x-vector}\cite{snyder2018x}}                 & \multicolumn{1}{c|}{0.56} & \multicolumn{1}{c|}{\textbf{1.43}} &\textbf{1.60 } \\ \hline
\multicolumn{1}{|c|}{\textbf{RawNet3} \cite{jung2022pushing}}                  & \multicolumn{1}{c|}{0.57} & \multicolumn{1}{c|}{1.44} & 1.75 \\ \hline
\multicolumn{1}{|c|}{\textbf{Verispeak} \cite{Verispeak}}                  & \multicolumn{1}{c|}{\textbf{0.35}} & \multicolumn{1}{c|}{1.75} & 1.83 \\ \hline

\end{tabular}

\label{tab:EER_MAVS}
\end{table}

\section{Vulnerability analysis}
\label{sec:vulnerability}
This section presents the vulnerability analysis of the proposed morphing on three SVS. The proposed morph samples are generated based on the gender pairs i,e., male-male, female-female and combined pairs. We benchmark the attack potential of TD-VIM by comparing the verification scores computed from both the contributory subjects from the above said pairs by setting False Alarm Rate (FAR)=0.1\%. We enroll the morph sample to the SVS which consists of speaker characteristics of both contributory subjects and while probing we probe the indivisual speech samples. If both the contributory subjects are verified while probing then only the morph is considered to be successful. We do this process for both text-dependent and text-independent cases (results are provided in the supplementary material). Before we proceed onto the vulnerability analysis we introduce the metrics used to calculate the vulnerability of SVS.

\subsection{Metrics used for calculating the vulnerability}
The vulnerability of the SVS can be calculated using three different types of metrices namely: Mated Morphed Presentation Match Rate (MMPMR) \cite{scherhag2017biometric}, Fully Mated Morphed Presentation Match Rate (FMMPMR) \cite{venkatesh2020influence}, Morphing Attack Potential (MAP) \cite{ferrara2022morphing}. The MMPMR metric is based on distinctive trials, whereas FMMPMR is based on discriminant (pairwise) probe attempts of the contributory pairs. The MAP revamps the existing metrics by indicating the vulnerability as a matrix using multiple SVS with discriminant (pairwise) probe attempts. The main limitation of the MAP is that (a) does not quantify the vulnerability as a single number, which makes it difficult to compare the attack potential of multiple recognition systems and morphing generation algorithms.  (2) does not consider Failure-to-Acquire Rate (FTAR) which is essential because the recognition algorithms are evaluated as a black box. These factors motivated us to employ the Generalised Morphing Attack Potential (G-MAP) \cite{singh2023deep} to benchmark the attack potential of the proposed method as it can (1) quantifies the vulnerability as a single number (2) takes into account of multiple morph generation types (3) consideration of FTAR.


\subsubsection*{Mathematical Expression for G-MAP}
Let $\mathbb{P}$ denotes the set of paired speech samples (in our case gender pair and also be denoted as number of probe attempts), let $\mathbb{S}$ denote the set of SVS, let $\mathbb{G}$ denote the morph attack generation algorithm, Let $\mathbb{M}_d$ denote morph speech set corresponding to $\mathbb{D}$, let $\tau_l$ similarity score from SVS (l), then G-MAP is defined as follows \cite{singh2023deep}:

\begin{equation}
\begin{aligned}
&{\textrm{G-MAP}} ={\frac{1}{|\mathbb{G}|}}{\sum_{d}^{|\mathbb{G}|}}{\frac{1}{|\mathbb{P}|}}{\frac{1}{|\mathbb{M}_d|}} \min_{\mathbb{F}_{l}}
& \sum_{i,j}^{|\mathbb{P}|,|\mathbb{M}_d|}\bigg\{\left[ (S1_i^j > \tau_l) \wedge \cdots ( Sk_i^j > \tau_l) \right] &{\times}  \left[ (1-FTAR(i,l)) \right] \bigg\}\\
\end{aligned}
\label{eq:gmap}
\end{equation}


where $FTAR(i,l)$ is the failure to acquire the probe speech sample in attempt i using SVS(l).

Since we have defined G-MAP in the Equation \ref{eq:gmap}, which includes multiple probe attempts, multiple SVS, and morph attack generation type. G-MAP with multiple probe attempts is calculated by setting $\mathbb{G}=1$ and $\mathbb{S}=1$ in the equation \ref{eq:gmap}. Now with FTAR=0, and similarity scores $S1_i^j$ os greater than threshold $\tau_l$, G-MAP with multiple probe attempts is equal to FMMPMR and G-MAP with multiple probe attempts and multiple SVS with $\mathbb{G}=1$ is computed by taking minimum of vulnerability obtained from multiple SVS. Now G-MAP would quantify vulnerability as single number whereas the MAP represents vulnerability as matrix of SVS. Finally G-MAP in which multiple attempts, multiple SVS, multiple attack types and FTAR provides vulnerability in terms of a single number.


\begin{figure*}[!h]
\centering
\includegraphics[width=0.75\textwidth]{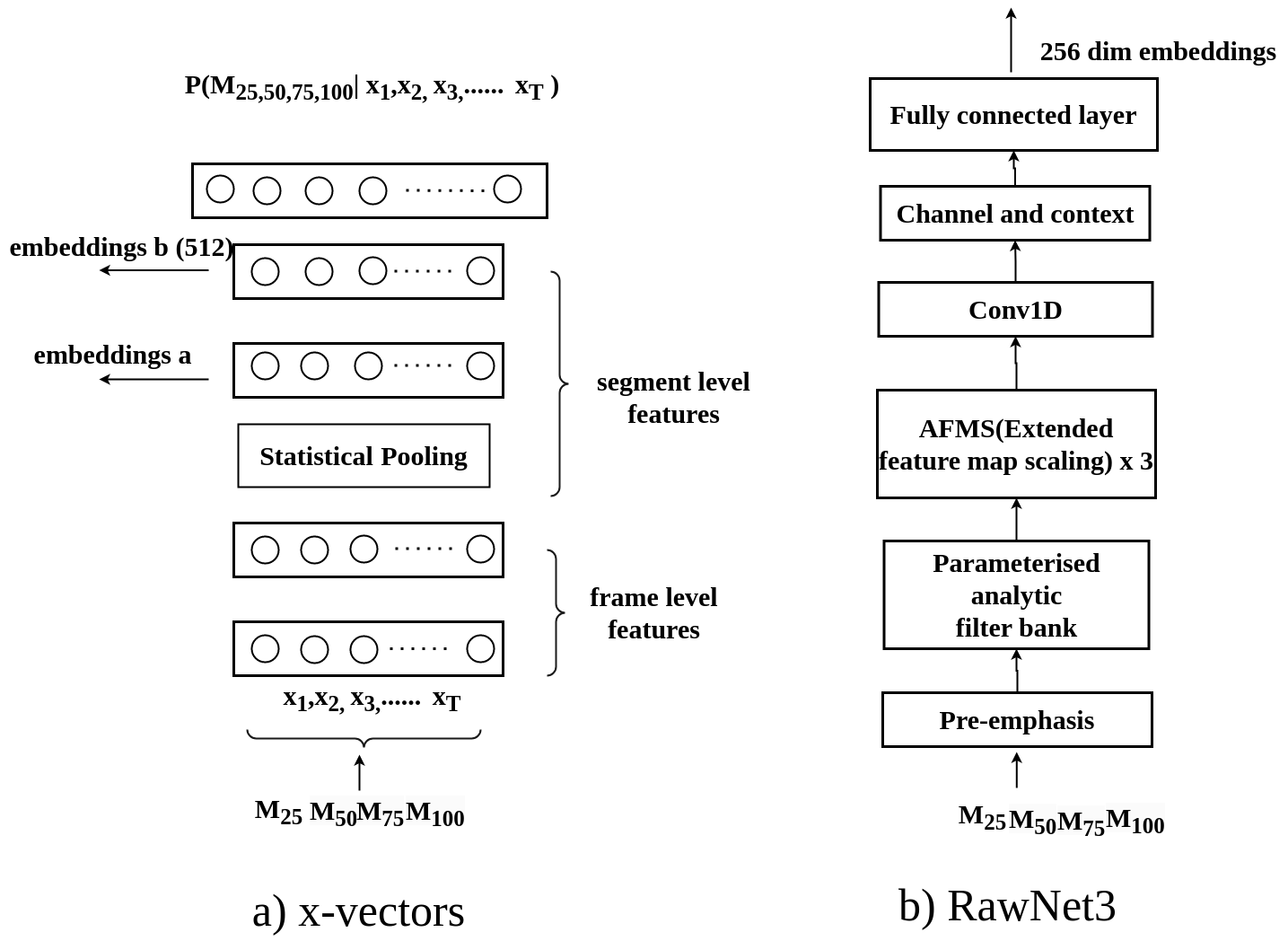}
\caption{The diagram showing $M_{25}$, $M_{50}$, $M_{75}$ and $M_{100}$ signal passed onto the x-vector and RawNet3 to get 512-dimensional and 256-dimensional embeddings respectively}
\label{fig:x_vec}
\end{figure*}

\subsection{Quantitative evaluation of Vulnerability using text-dependent morphing}
In text-dependent morphing, we enroll the morph sample (for example created from sentence S1 using four different morphing types) and while probing both the contributory subjects have to utter the same sentence (for example S1).
In the following sections,  we present the quantitative evaluation of the vulnerability corresponding to three SVS, x-vector, RawNet3 and verispeak for four different types of morphing. Because G-MAP is based on the number of shots, Multiple SVS, and morphing types (in our case it is 4), this will permit one to analyze the results based on a) probe attempts independent of SVS and attack speech generation category,  b) multiple SVS with multiple attempts, and c) G-MAP with a function of number of shots, multiple SVS, and corresponding different categories of speech attack generation type along with FTAR. However the VIM \cite{pani2023voice} models are not available in public, upon requesting multiple times also there was no response from authors.

\begin{table*}[!h]
\caption{Vulnerability analysis for MAVS database using G-MAP with multiple probe attempts on x-vector}
\footnotesize
\centering
\begin{tabular}{|c|c|cccccc|}
\hline
\multirow{4}{*}{\textbf{Morphing Factor}} & \multirow{4}{*}{\textbf{Language}} & \multicolumn{6}{c|}{\textbf{G-MAP with Multiple probe attempts}}                                                                                                      \\ \cline{3-8} 
                               &                           & \multicolumn{3}{c|}{\textbf{iPhone11}}                                                           & \multicolumn{3}{c|}{\textbf{Samsung S8}}                                    \\ \cline{3-8} 
                               &                           & \multicolumn{3}{c|}{\textbf{Gender Pair}}                                                        & \multicolumn{3}{c|}{\textbf{Gender Pair}}                                   \\ \cline{3-8} 
                               &                           & \multicolumn{1}{c|}{\textbf{FF}}    & \multicolumn{1}{c|}{\textbf{MM}}    & \multicolumn{1}{c|}{\textbf{Combined}} & \multicolumn{1}{c|}{\textbf{FF}}    & \multicolumn{1}{c|}{\textbf{MM}}    & \textbf{Combined} \\ \hline
\multirow{3}{*}{$M_{25}$}           & English                   & \multicolumn{1}{c|}{23.76} & \multicolumn{1}{c|}{24.74} & \multicolumn{1}{c|}{48.54}    & \multicolumn{1}{c|}{67.36} & \multicolumn{1}{c|}{68.16} & 67.76    \\ \cline{2-8} 
                               & Hindi                     & \multicolumn{1}{c|}{23.78} & \multicolumn{1}{c|}{24.80} & \multicolumn{1}{c|}{48.58}    & \multicolumn{1}{c|}{69.73} & \multicolumn{1}{c|}{69.36} & 69.54    \\ \cline{2-8} 
                               & Bengali                   & \multicolumn{1}{c|}{22.2} & \multicolumn{1}{c|}{23.69} & \multicolumn{1}{c|}{45.89}    & \multicolumn{1}{c|}{67.37} & \multicolumn{1}{c|}{68.26} & 67.81    \\ \hline
\multirow{3}{*}{$M_{50}$}           & English                   & \multicolumn{1}{c|}{25.79} & \multicolumn{1}{c|}{26.78} & \multicolumn{1}{c|}{52.57}     & \multicolumn{1}{c|}{68.34} & \multicolumn{1}{c|}{69.87} & 69.10    \\ \cline{2-8} 
                               & Hindi                     & \multicolumn{1}{c|}{24.67} & \multicolumn{1}{c|}{25.69} & \multicolumn{1}{c|}{50.36}    & \multicolumn{1}{c|}{69.44} & \multicolumn{1}{c|}{69.35} & 69.39    \\ \cline{2-8} 
                               & Bengali                   & \multicolumn{1}{c|}{23.59} & \multicolumn{1}{c|}{25.45} & \multicolumn{1}{c|}{49.04}     & \multicolumn{1}{c|}{69.87} & \multicolumn{1}{c|}{68.36} & 69.11    \\ \hline
\multirow{3}{*}{$M_{75}$}           & English                   & \multicolumn{1}{c|}{27.17} & \multicolumn{1}{c|}{28.62} & \multicolumn{1}{c|}{52.57}    & \multicolumn{1}{c|}{68.19} & \multicolumn{1}{c|}{68.25} & 68.57    \\ \cline{2-8} 
                               & Hindi                     & \multicolumn{1}{c|}{26.79} & \multicolumn{1}{c|}{26.58} & \multicolumn{1}{c|}{53.37}     & \multicolumn{1}{c|}{68.69} & \multicolumn{1}{c|}{68.45} & 68.57    \\ \cline{2-8} 
                               & Bengali                   & \multicolumn{1}{c|}{24.02} & \multicolumn{1}{c|}{26.58} & \multicolumn{1}{c|}{50.6}    & \multicolumn{1}{c|}{69.2}  & \multicolumn{1}{c|}{69.12} & 69.16    \\ \hline
\multirow{3}{*}{$M_{100}$}          & English                   & \multicolumn{1}{c|}{29.12} & \multicolumn{1}{c|}{29.34} & \multicolumn{1}{c|}{58.46}    & \multicolumn{1}{c|}{56.14} & \multicolumn{1}{c|}{56.74} & 56.44    \\ \cline{2-8} 
                               & Hindi                     & \multicolumn{1}{c|}{27.79} & \multicolumn{1}{c|}{28.67} & \multicolumn{1}{c|}{56.34}    & \multicolumn{1}{c|}{57.64} & \multicolumn{1}{c|}{57.99} & 57.81    \\ \cline{2-8} 
                               & Bengali                   & \multicolumn{1}{c|}{25.78} & \multicolumn{1}{c|}{27.96} & \multicolumn{1}{c|}{53.71}    & \multicolumn{1}{c|}{59.85} & \multicolumn{1}{c|}{59.14} & 59.49    \\ \hline
\end{tabular}
\label{tab:G-map_MAVS_MA_xvec}
\end{table*}

\begin{table*}[!h]
\caption{Vulnerability analysis for MAVS database using G-MAP with multiple probe attempts on RawNet3}
\footnotesize
\centering
\begin{tabular}{|c|c|cccccc|}
\hline
\multirow{4}{*}{\textbf{\makecell{Morphing Factor}}} & \multirow{4}{*}{\textbf{\makecell{Language}}} & \multicolumn{6}{c|}{\textbf{\makecell{G-MAP with Multiple probe attempts}}}                                                                                                      \\ \cline{3-8} 
                               &                           & \multicolumn{3}{c|}{\textbf{\makecell{iPhone11}}}                                                           & \multicolumn{3}{c|}{\textbf{\makecell{Samsung S8}}}                                    \\ \cline{3-8} 
                               &                           & \multicolumn{3}{c|}{\textbf{\makecell{Gender Pair}}}                                                        & \multicolumn{3}{c|}{\textbf{\makecell{Gender Pair}}}                                   \\ \cline{3-8} 
                               &                           & \multicolumn{1}{c|}{\textbf{\makecell{FF}}}    & \multicolumn{1}{c|}{\textbf{\makecell{MM}}}    & \multicolumn{1}{c|}{\textbf{\makecell{Combined}}} & \multicolumn{1}{c|}{\textbf{\makecell{FF}}}    & \multicolumn{1}{c|}{\textbf{\makecell{MM}}}    & {\textbf{\makecell{Combined}}} \\ \hline
\multirow{3}{*}{$M_{25}$}           & English                   & \multicolumn{1}{c|}{97.21} & \multicolumn{1}{c|}{91.09} & \multicolumn{1}{c|}{94.15}    & \multicolumn{1}{c|}{97.52} & \multicolumn{1}{c|}{96.30} & 96.91    \\ \cline{2-8} 
                               & Hindi                     & \multicolumn{1}{c|}{98.11} & \multicolumn{1}{c|}{90.68} & \multicolumn{1}{c|}{94.39}    & \multicolumn{1}{c|}{98.78} & \multicolumn{1}{c|}{97.65} & 98.21    \\ \cline{2-8} 
                               & Bengali                   & \multicolumn{1}{c|}{99.12} & \multicolumn{1}{c|}{91.58} & \multicolumn{1}{c|}{95.35}    & \multicolumn{1}{c|}{99.68} & \multicolumn{1}{c|}{98.87} & 99.27    \\ \hline
\multirow{3}{*}{$M_{50}$}           & English                   & \multicolumn{1}{c|}{96.09} & \multicolumn{1}{c|}{87.48} & \multicolumn{1}{c|}{91.78}    & \multicolumn{1}{c|}{96.25} & \multicolumn{1}{c|}{97.5}  & 96.87    \\ \cline{2-8} 
                               & Hindi                     & \multicolumn{1}{c|}{97.29} & \multicolumn{1}{c|}{88.68} & \multicolumn{1}{c|}{92.96}    & \multicolumn{1}{c|}{97.85} & \multicolumn{1}{c|}{98.74} & 98.29    \\ \cline{2-8} 
                               & Bengali                   & \multicolumn{1}{c|}{98.45} & \multicolumn{1}{c|}{89.18} & \multicolumn{1}{c|}{93.81}    & \multicolumn{1}{c|}{98.25} & \multicolumn{1}{c|}{99.41} & 98.83    \\ \hline
\multirow{3}{*}{$M_{75}$}           & English                   & \multicolumn{1}{c|}{97.4}  & \multicolumn{1}{c|}{88.56} & \multicolumn{1}{c|}{92.98}    & \multicolumn{1}{c|}{97.21} & \multicolumn{1}{c|}{96.87} & 97.04    \\ \cline{2-8} 
                               & Hindi                     & \multicolumn{1}{c|}{98.04} & \multicolumn{1}{c|}{89.77} & \multicolumn{1}{c|}{93.05}    & \multicolumn{1}{c|}{98.12} & \multicolumn{1}{c|}{97.36} & 97.72    \\ \cline{2-8} 
                               & Bengali                   & \multicolumn{1}{c|}{99.40} & \multicolumn{1}{c|}{90.67} & \multicolumn{1}{c|}{95.03}    & \multicolumn{1}{c|}{99.21} & \multicolumn{1}{c|}{98.76} & 98.98    \\ \hline
\multirow{3}{*}{$M_{100}$}          & English                   & \multicolumn{1}{c|}{97.1}  & \multicolumn{1}{c|}{92.94} & \multicolumn{1}{c|}{95.02}    & \multicolumn{1}{c|}{96.26} & \multicolumn{1}{c|}{97.14} & 96.7     \\ \cline{2-8} 
                               & Hindi                     & \multicolumn{1}{c|}{98.67} & \multicolumn{1}{c|}{93.8}  & \multicolumn{1}{c|}{96.23}    & \multicolumn{1}{c|}{97.36} & \multicolumn{1}{c|}{98.74} & 98.10    \\ \cline{2-8} 
                               & Bengali                   & \multicolumn{1}{c|}{99.57} & \multicolumn{1}{c|}{94.82} & \multicolumn{1}{c|}{97.19}    & \multicolumn{1}{c|}{98.26} & \multicolumn{1}{c|}{99.74} & 99.01    \\ \hline
\end{tabular}
\label{tab:G-map_MAVS_MA_rawnet}
\end{table*}

\subsubsection{Vulnerability of x-vectors for MAVS Database}
To test the effectiveness of G-MAP on different devices and SVS, we selected two latest phone models from the MAVS database: iPhone 11 and Samsung S8. We evaluated the performance of G-MAP on each device, conducted multiple probe attempts, and compared its results with three different SVS systems. This setup allowed us to analyze the accuracy and consistency of G-MAP across different devices and SVS implementations. Table \ref{tab:G-map_MAVS_MA_xvec} lists G-MAP (multiple probe attempts in our case it is set to 3) for the x-vector. We enroll the morph sample, and during probing, we examine both the contributory subjects. For the morph sample, we extract 512-dimensional embeddings, and similarly, we extract 512-dimensional embeddings for the contributory subjects. The cosine similarity between the embeddings of the morphed sample and both contributory subjects is calculated. If the cosine similarity exceeds a predefined threshold for both contributory subjects, only then is the attack considered successful. For x-vectors the operation threshold is set at 0.5. From the table, the following points can be observed:

\begin{table*}[!]
\caption{Vulnerability analysis for MAVS database using G-MAP with multiple probe attempts for the Verispeak commercial speaker verification system}
\footnotesize
\centering
\begin{tabular}{|c|c|cccccc|}
\hline
\multirow{4}{*}{\textbf{\makecell{Morphing Factor}}} & \multirow{4}{*}{\textbf{\makecell{Language}}} & \multicolumn{6}{c|}{\textbf{\makecell{G-MAP with Multiple probe attempts}}}                                                                                                      \\ \cline{3-8} 
                               &                           & \multicolumn{3}{c|}{\textbf{\makecell{iPhone11}}}                                                           & \multicolumn{3}{c|}{\textbf{\makecell{Samsung S8}}}                                    \\ \cline{3-8} 
                               &                           & \multicolumn{3}{c|}{\textbf{\makecell{Gender Pair}}}                                                        & \multicolumn{3}{c|}{\textbf{\makecell{Gender Pair}}}                                   \\ \cline{3-8} 
                               &                           & \multicolumn{1}{c|}{\textbf{\makecell{FF}}}    & \multicolumn{1}{c|}{\textbf{\makecell{MM}}}    & \multicolumn{1}{c|}{\textbf{\makecell{Combined}}} & \multicolumn{1}{c|}{\textbf{\makecell{FF}}}    & \multicolumn{1}{c|}{\textbf{\makecell{MM}}}    & {\textbf{\makecell{Combined}}} \\ \hline
\multirow{3}{*}{M25}           & English                   & \multicolumn{1}{c|}{50.21} & \multicolumn{1}{c|}{51.09} & \multicolumn{1}{c|}{50.65}    & \multicolumn{1}{c|}{51.52} & \multicolumn{1}{c|}{51.30} & 51.41    \\ \cline{2-8} 
                               & Hindi                     & \multicolumn{1}{c|}{53.11} & \multicolumn{1}{c|}{53.68} & \multicolumn{1}{c|}{53.39}    & \multicolumn{1}{c|}{52.78} & \multicolumn{1}{c|}{52.65} & 52.71    \\ \cline{2-8} 
                               & Bengali                   & \multicolumn{1}{c|}{51.12} & \multicolumn{1}{c|}{50.58} & \multicolumn{1}{c|}{50.84}    & \multicolumn{1}{c|}{50.68} & \multicolumn{1}{c|}{52.87} & 51.77    \\ \hline
\multirow{3}{*}{M50}           & English                   & \multicolumn{1}{c|}{55.09} & \multicolumn{1}{c|}{54.48} & \multicolumn{1}{c|}{54.78}    & \multicolumn{1}{c|}{56.25} & \multicolumn{1}{c|}{54.5}  & 55.37    \\ \cline{2-8} 
                               & Hindi                     & \multicolumn{1}{c|}{56.29} & \multicolumn{1}{c|}{55.68} & \multicolumn{1}{c|}{55.98}    & \multicolumn{1}{c|}{57.85} & \multicolumn{1}{c|}{57.74} & 57.79    \\ \cline{2-8} 
                               & Bengali                   & \multicolumn{1}{c|}{54.45} & \multicolumn{1}{c|}{54.18} & \multicolumn{1}{c|}{54.36}    & \multicolumn{1}{c|}{56.25} & \multicolumn{1}{c|}{55.41} & 55.83    \\ \hline
\multirow{3}{*}{M75}           & English                   & \multicolumn{1}{c|}{59.4}  & \multicolumn{1}{c|}{58.56} & \multicolumn{1}{c|}{54.36}    & \multicolumn{1}{c|}{60.21} & \multicolumn{1}{c|}{59.87} & 60.04    \\ \cline{2-8} 
                               & Hindi                     & \multicolumn{1}{c|}{62.04} & \multicolumn{1}{c|}{61.77} & \multicolumn{1}{c|}{61.90}    & \multicolumn{1}{c|}{61.12} & \multicolumn{1}{c|}{60.36} & 60.73    \\ \cline{2-8} 
                               & Bengali                   & \multicolumn{1}{c|}{61.90} & \multicolumn{1}{c|}{61.67} & \multicolumn{1}{c|}{62.53}    & \multicolumn{1}{c|}{62.21} & \multicolumn{1}{c|}{61.76} & 61.98    \\ \hline
\multirow{3}{*}{M100}          & English                   & \multicolumn{1}{c|}{70.1}  & \multicolumn{1}{c|}{69.94} & \multicolumn{1}{c|}{70.02}    & \multicolumn{1}{c|}{69.26} & \multicolumn{1}{c|}{68.14} & 68.70     \\ \cline{2-8} 
                               & Hindi                     & \multicolumn{1}{c|}{68.67} & \multicolumn{1}{c|}{67.8}  & \multicolumn{1}{c|}{68.23}    & \multicolumn{1}{c|}{69.36} & \multicolumn{1}{c|}{68.74} & 69.05    \\ \cline{2-8} 
                               & Bengali                   & \multicolumn{1}{c|}{70.57} & \multicolumn{1}{c|}{67.82} & \multicolumn{1}{c|}{69.15}    & \multicolumn{1}{c|}{69.26} & \multicolumn{1}{c|}{66.74} & 68    \\ \hline
\end{tabular}
\label{tab:G-map_verispeak_mavs}
\end{table*}

\begin{itemize}
    \item From the Table \ref{tab:G-map_MAVS_MA_xvec} we can see a great extent of device dependency where iPhone 11 is less vulnerable and Samsung S8 phones are most vulnerable for x-vectors irrespective of the language.
    \begin{itemize}
        \item iPhone 11 shows lower vulnerability to the morphing attacks because iPhone often employs advanced microphone arrays, making it harder for morphed signals to fool the system. 
        \item On the other hand, it is possible that Samsung S8 smartphones are more susceptible to attacks due to differences in hardware configurations, signal processing mechanisms, or the complexity of their audio processing capabilities. It is important to note that the specific vulnerabilities of a given device may depend on a variety of factors, and a comprehensive assessment should take into account the unique characteristics of each device.
    \end{itemize}

     \item In the case of the iPhone-11 the $M_{25}$ component utilizes only a portion of the speech sample, specifically only the initial frames, to extract the second speaker's characteristics. These characteristics are then passed through a time-delayed neural network (TDNN) to generate frame-level embeddings. The frames were then statistically pooled to create a single vector, which was subsequently aggregated into segment-level features, as illustrated in Figure \ref{fig:x_vec}. As a result, this type of morphing demonstrated lower vulnerability.
    \item The $M_{50}$ method, in contrast to the $M_{25}$ technique, is marginally more susceptible to morphing attacks. This is due to the fact that the first half of the speech samples is altered in $M_{50}$, which includes characteristics of the second speaker, thereby significantly enhancing the impact of morphing on the overall representation of the speakers. Despite this, the x-vector architecture continues to employ statistical pooling to combine features, as illustrated in Figure \ref{fig:x_vec}, which may lead to the verification of both individuals.

\item The vulnerability of x-vector-based SVS increases as the proportion of morphed speech containing the second speaker's characteristics grows. In the $M_{75}$ type, where three-fourth of the morphed signal comprises the second speaker's voice, as shown in Figure \ref{fig:x_vec}, the SVS exhibits a higher vulnerability than both $M_{25}$ and $M_{50}$. This increased vulnerability arises because a larger portion of morphed frames exerts a stronger influence on the overall speaker representation during statistical pooling. Consequently, the genuine speaker's characteristics are overshadowed, leading to a higher risk of  false verification by SVS.

\item In the $M_{100}$ morphing type, where the morphed speech represents the average of both contributing speakers' voices, the susceptibility of x-vector-based SVS reaches its peak. This morphing method seamlessly merges the traits of both speakers across the entire signal, making it challenging for the SVS to differentiate between genuine speakers and verify both speakers.

    \item It is noteworthy that, in the context of x-vector-based SVS, the $M_{25}$ morph exhibits greater susceptibility compared to the $M_{100}$ morph, especially on Samsung S8 devices. This unanticipated behavior may be attributable to various factors, including the x-vector model's responsiveness to certain levels or forms of voice manipulation.
    
    \item This unanticipated behavior underscores the intricate connection between the subtleties of varying morphing levels and the particular responses of SVS models across different devices. Further investigation is required to elucidate why a less intensive morphing method, such as $M_{25}$, leads to increased vulnerability when contrasted with a more extensive one, like $M_{100}$, in the context of Samsung S8 smartphones and x-vector-based SVS models.

\end{itemize}

\begin{table*}[!]
\caption{Vulnerability analysis for MAVS database using G-MAP with multiple probe attempts with multiple SVS}
\footnotesize
\centering
\begin{tabular}{|c|c|cccccc|}
\hline
\multirow{4}{*}{\textbf{Morphing factor}} & \multirow{4}{*}{\textbf{Language}} & \multicolumn{6}{c|}{\textbf{G-MAP with Multiple probe attempts with multiple SVS}}                                                                                                      \\ \cline{3-8} 
                               &                           & \multicolumn{3}{c|}{\textbf{iPhone11}}                                                           & \multicolumn{3}{c|}{\textbf{Samsung S8}}                                    \\ \cline{3-8} 
                               &                           & \multicolumn{3}{c|}{\textbf{Gender Pair}}                                                        & \multicolumn{3}{c|}{\textbf{Gender Pair}}                                   \\ \cline{3-8} 
                               &                           & \multicolumn{1}{c|}{\textbf{FF}}    & \multicolumn{1}{c|}{\textbf{MM}}    & \multicolumn{1}{c|}{\textbf{Combined}} & \multicolumn{1}{c|}{\textbf{FF}}    & \multicolumn{1}{c|}{\textbf{MM}}    & \textbf{Combined} \\ \hline
\multirow{3}{*}{$M_{25}$}           & English                   & \multicolumn{1}{c|}{23.76} & \multicolumn{1}{c|}{24.74} & \multicolumn{1}{c|}{48.54}    & \multicolumn{1}{c|}{51.52} & \multicolumn{1}{c|}{51.30} & 51.41    \\ \cline{2-8} 
                               & Hindi                     & \multicolumn{1}{c|}{23.78} & \multicolumn{1}{c|}{24.80} & \multicolumn{1}{c|}{48.58}    & \multicolumn{1}{c|}{52.78} & \multicolumn{1}{c|}{52.65} & 52.71    \\ \cline{2-8} 
                               & Bengali                   & \multicolumn{1}{c|}{22.2} & \multicolumn{1}{c|}{23.69} & \multicolumn{1}{c|}{45.89}    & \multicolumn{1}{c|}{50.68} & \multicolumn{1}{c|}{52.87} & 51.77    \\ \hline
\multirow{3}{*}{$M_{50}$}           & English                   & \multicolumn{1}{c|}{25.79} & \multicolumn{1}{c|}{26.78} & \multicolumn{1}{c|}{52.57}     & \multicolumn{1}{c|}{56.25} & \multicolumn{1}{c|}{54.5} & 55.37    \\ \cline{2-8} 
                               & Hindi                     & \multicolumn{1}{c|}{24.69} & \multicolumn{1}{c|}{25.69} & \multicolumn{1}{c|}{50.36}    & \multicolumn{1}{c|}{57.85} & \multicolumn{1}{c|}{57.74} & 55.79    \\ \cline{2-8} 
                               & Bengali                   & \multicolumn{1}{c|}{23.59} & \multicolumn{1}{c|}{25.45} & \multicolumn{1}{c|}{49.04}     & \multicolumn{1}{c|}{56.25} & \multicolumn{1}{c|}{55.41} & 55.83    \\ \hline
\multirow{3}{*}{$M_{75}$}           & English                   & \multicolumn{1}{c|}{27.17} & \multicolumn{1}{c|}{28.62} & \multicolumn{1}{c|}{52.57}    & \multicolumn{1}{c|}{60.21} & \multicolumn{1}{c|}{59.87} & 60.04    \\ \cline{2-8} 
                               & Hindi                     & \multicolumn{1}{c|}{26.79} & \multicolumn{1}{c|}{26.58} & \multicolumn{1}{c|}{53.37}     & \multicolumn{1}{c|}{61.12} & \multicolumn{1}{c|}{60.36} & 60.73    \\ \cline{2-8} 
                               & Bengali                   & \multicolumn{1}{c|}{24.02} & \multicolumn{1}{c|}{26.58} & \multicolumn{1}{c|}{50.6}    & \multicolumn{1}{c|}{62.21}  & \multicolumn{1}{c|}{61.76} & 61.98    \\ \hline
\multirow{3}{*}{$M_{100}$}          & English                   & \multicolumn{1}{c|}{29.12} & \multicolumn{1}{c|}{29.34} & \multicolumn{1}{c|}{58.46}    & \multicolumn{1}{c|}{56.14} & \multicolumn{1}{c|}{56.74} & 56.44    \\ \cline{2-8} 
                               & Hindi                     & \multicolumn{1}{c|}{27.79} & \multicolumn{1}{c|}{28.67} & \multicolumn{1}{c|}{56.34}    & \multicolumn{1}{c|}{57.64} & \multicolumn{1}{c|}{57.99} & 57.81    \\ \cline{2-8} 
                               & Bengali                   & \multicolumn{1}{c|}{25.78} & \multicolumn{1}{c|}{27.96} & \multicolumn{1}{c|}{53.71}    & \multicolumn{1}{c|}{59.85} & \multicolumn{1}{c|}{59.14} & 59.49    \\ \hline
\end{tabular}
\label{tab:G-map_MAVS_MA_vrs}
\end{table*}

\subsubsection{Vulnerability of RawNet3 for MAVS Database}
Similar to x-vectors, we calculate the 256-dimensional embeddings for the morph signal and for both contributory signals and then calculate the cosine similarity between the morph signal and two contributory signals. If the cosine similarity is greater than a pre-defined threshold then the attack is considered as succesful. For RawNet3 we set a threshold of 0.8.
Table \ref{tab:G-map_MAVS_MA_rawnet} shows the quantitative analysis (G-map with multiple probe attempts) for the MAVS database. The following points can be observed from the Table \ref{tab:G-map_MAVS_MA_rawnet}:

\begin{itemize}
    \item It is uncommon to witness complete vulnerability in the $M_{25}$ morphing type, given that only the initial signal possesses the characteristics of the second speaker. The pre-emphasis technique enhances the high-frequency components of a speech signal relative to the low-frequency components. 
    \item In the next stage the pre-emphasised signal is passed onto the parameterised analytic filter bank. The analytic filterbank produces 10 filterbank outputs. The more details regarding the visualisation of pre-emphasis signals for both morph and original and visualisation of filterbank outputs are shown in Supplementary material.

\item Three backbone blocks digest these filterbank output. Each backbone block is referred to as AFMS-Res2MP block. The AFMS (Extended Feature Map Scaling) module performs operations like scaling, normalization, enhancement of feature maps. This scaling process  involve various techniques aimed at improving the representation or characteristics of the extracted features.

\item The SVS based on RawNet3 has demonstrated consistent susceptibility to morphing attacks, regardless of the device type, whether it's an iPhone-11 or a Samsung S8. The fact that this susceptibility is consistent across all morphing types suggests that the vulnerability of RawNet3 to morphing attacks is unaffected by the specific device being used.

\item Despite the differences in hardware features, configurations, and signal processing capabilities between iPhone 11 and Samsung S8 devices, RawNet3-based SVS has consistently failed to meet the demands posed by a variety of morphing techniques.

\item Our newly proposed morphing method has proven successful in deceiving the RawNet3-based SVS. This was accomplished through the utilization of various morphing techniques, like $M_{25}$, $M_{50}$, $M_{75}$, and $M_{100}$, which were applied to all three languages. The results of our study have exposed vulnerabilities in the SVS, demonstrating the effectiveness of our technique.

\end{itemize}

\subsubsection{Vulnerability analysis of  Verispeak for MAVS database}
Verispeak \cite{Verispeak} a CTOS system is used for speaker verification.The operating principle of SVS is unkown and not disclosed. We evaluated the performance of our morphing process by testing a morphed sample against both contributing subjects at a false match rate (FMR) of 0.1\%. Despite the effectiveness of our approach,  Table \ref{tab:G-map_verispeak_mavs} provides a comprehensive overview of our results, showcasing the G-MAP with multiple probe attempts for the proposed morphing algorithm. Upon examining the table, several key observations can be made.

\begin{itemize}
     \item Upon analyzing the Verispeak based SVS for the $M_{25}$ morphing type, it became evident that the vulnerability of the model was significantly reduced. This reduction in vulnerability can be attributed to the fact that only one segment of the speech sample exhibited the characteristics of a second speaker.
    \item With the increasing influence of the second speaker's characteristics, it can be observed that the vulnerability of $M_{50}$ is slightly greater than that of $M_{25}$.
    \item The susceptibility to influence is heightened with morphing methods, such as $M_{75}$, which prominently incorporate the characteristics of the second speaker.
    \item In instances involving the $M_{100}$ type, the morphed speech represents the average characteristics of both contributing speakers voices. The vulnerability of the Verispeak SVS increases in such cases. This morphing method effectively combines the features of both speakers throughout the entire signal, presenting significant obstacles for Verispeak in accurately distinguishing genuine speakers.
    \item The results demonstrate that the morphing technique exhibits a higher susceptibility towards Verispeak, a commercial biometric speaker verification system, when compared with x-vector-based SVS.
    \item For Verispeak SVS also we can see great amount of device dependency. The Samsung phones shows higher vulnerability while iPhone-11 shows lower vulnerability again this can be attributed to the fact that differences in the microphone arrays, signal quality, audio-processing mechanisms employed by the different manufacturer.
    \end{itemize}
\begin{table}[!h]
\centering
\caption{G-MAP (with full capacity) for iPhone-11 and Samsung S8}
\label{tab:final_gmap_td}
\begin{tabular}{|cc|}
\hline
\multicolumn{2}{|c|}{\textbf{G-MAP(\%)}}    \\ \hline
\multicolumn{1}{|c|}{\textbf{iPhone-11}} &\textbf{Samsung S8}  \\ \hline
\multicolumn{1}{|c|}{52.08} &56.61  \\ \hline
\end{tabular}
\end{table}

A morphed sample is considered vulnerable if multiple probes attempt to successfully deceive multiple SVS. Therefore, G-MAP offers a singular value indicating vulnerability by averaging probe attempts while considering the FTAR. Table \ref{tab:G-map_MAVS_MA_vrs} indicates the G-MAP (multiple probe attempts and multiple SVS) for MAVS database for three different languages. Based on the obtained results, the following observations were made:
\begin{itemize}
    \item G-MAP score with multiple probe attempts and multiple SVS is calculated by taking minimum of across all SVS and using $\mathbb{G}=1$ as per equation \ref{eq:gmap}.
    \item The iPhone-11 shows lower vulnerability than the Samsung-S8 phones due to the differences in hardware configurations, signal processing mechanisms.
\end{itemize}

Table \ref{tab:final_gmap_td} shows the vulnerability computed with the full capacity of G-MAP while considering multiple probe attempts, multiple SVS, multiple attack types, and FTAR. The G-MAP shown in Table \ref{tab:final_gmap_td} quantifies the vulnerability as a single number for each device.

\subsubsection{Results analysis}

\begin{figure*}[!h]
\begin{minipage}[c]{0.5\linewidth}
\includegraphics[width=\linewidth]{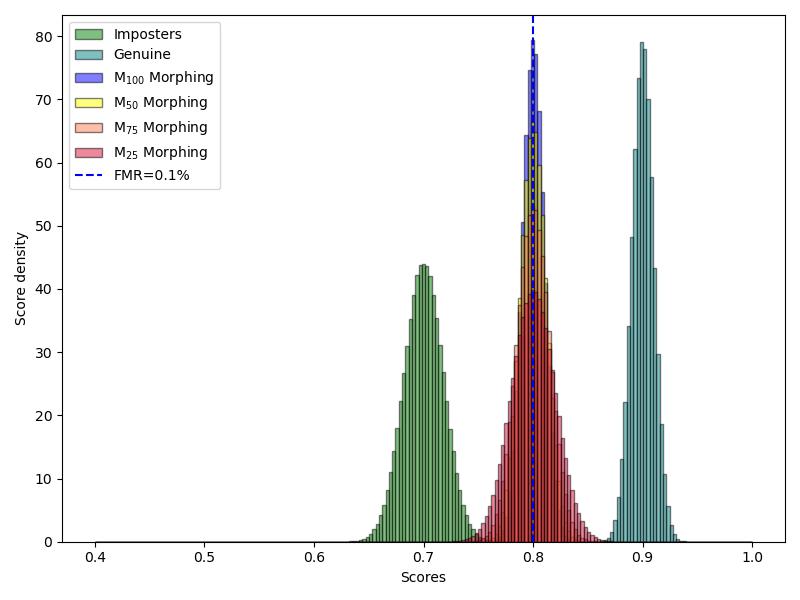}
\caption*{(a) Rawnet (TD)}
\end{minipage}
\hfill
\begin{minipage}[c]{0.5\linewidth}
\includegraphics[width=\linewidth]{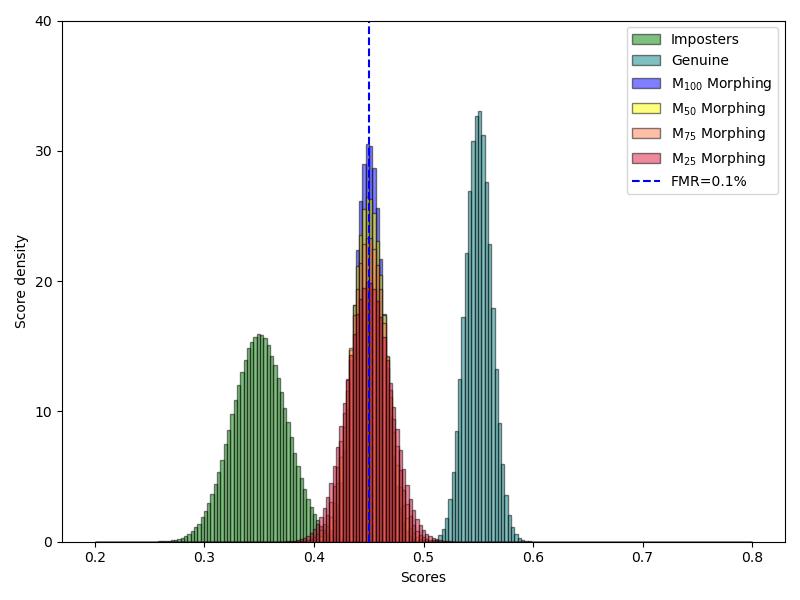}
\caption*{(b) x-vector (TI)}
\end{minipage}%
\caption{The speaker verification match score distribution comprises several comparisons: non-morph genuine pairs versus non-morph impostor pairs, as well as morph (across all four types) versus non-morph genuine pairs. This analysis employs Rawnet on iPhone-11 for English language and x-vector (on iPhone-11 for Hindi language).}
\label{fig:histogram_plots}
\end{figure*}

Further analysis of the morph attack performance is conducted using histogram plots. The histogram plots described in Figure \ref{fig:histogram_plots} depict the distribution of match scores for three different types of pairs in two speaker verification systems: genuine pairs (teal), impostor pairs (green), and pairs containing at least one morphed sample for four different morph types (crimson, yellow, coral and blue) for RawNet  and for x-vectors cases. From the histogram we observe the following points:
\begin{itemize}
    \item Genuine Pairs (Teal):
    \begin{itemize}
        \item Genuine pairs represent pairs of samples where both samples come from the same speaker.
        \item The match scores for genuine pairs typically have higher values since they represent a match between samples from the same speaker.
        \item In the histogram, the teal colour distribution is shifted towards higher match scores.
    \end{itemize}
    \item Impostor Pairs (Green):
    \begin{itemize}
        \item Impostor pairs consist of pairs of samples where each sample comes from a different speaker.
        \item The match scores for impostor pairs tend to be lower since there is no true match between the samples.
        \item The green distribution in the histogram is skewed towards lower match scores.
    \end{itemize}
    \item Pairs with at least one Morphed Sample (crimson, yellow, coral and blue):
    \begin{itemize}
        \item These pairs involve at least one sample that has been morphed or manipulated to resemble another speaker.
        \item The match scores for these pairs will be in between those of genuine and impostor pairs. They will be higher than those of regular impostor pairs but lower than those of genuine pairs.
        \item The crimson, yellow, coral and blue distribution in the histogram lies between the green and light blue distributions, indicating intermediate match scores.
    \end{itemize}
\end{itemize}

\section{Data Availability: }
The morphed files and original MAVS dataset is available from the corresponding author on reasonable request and to promote reproducibility and transparency, we provide the source code for TD-VIM in the repository https://github.com/Aravinda27/TD-VIM.

\section{Conclusion}
\label{sec:conclusion}

This work introduces Time-Domain Voice Identity Morphing (TD-VIM), a new technique for generating morphed voice signals directly at the signal level. Using the MAVS database, TD-VIM creates blended voice samples that combine characteristics from two distinct identities. We conducted an in-depth vulnerability assessment on several Speaker Verification Systems (SVS), including the x-vector, RawNet3, and a commercial system, Verispeak. Our evaluation used four distinct morphing attack types and a unique metric, Generalized Morph Attack Potential (G-MAP), to quantify vulnerabilities. The results reveal critical insights into how language and device types impact SVS susceptibility to morphing.

This work further explores G-MAP’s robustness across devices, affirming it as a versatile method for measuring SVS vulnerabilities. Our targeted analysis on Verispeak highlights TD-VIM's success rate in challenging advanced SVS defenses. The findings underscore TD-VIM’s potential to bypass sophisticated verification measures, emphasizing the importance of enhancing SVS security. 






 





\bibliography{sample}

@misc{bworld,
  author = {HSBC},
  title = {HSBC Voice ID},
  howpublished = {\url{https://hsbc.com.hk}},
  year = {2024}, 
  note = {[Online; Feb. 2024]}
}

@article{wu2015spoofing,
  title={Spoofing and countermeasures for speaker verification: A survey},
  author={Wu, Zhizheng and Evans, Nicholas and Kinnunen, Tomi and Yamagishi, Junichi and Alegre, Federico and Li, Haizhou},
  journal={speech communication},
  volume={66},
  pages={130--153},
  year={2015},
  publisher={Elsevier}
}

@article{khan2023battling,
  title={Battling voice spoofing: a review, comparative analysis, and generalizability evaluation of state-of-the-art voice spoofing counter measures},
  author={Khan, Awais and Malik, Khalid Mahmood and Ryan, James and Saravanan, Mikul},
  journal={Artificial Intelligence Review},
  volume={56},
  number={Suppl 1},
  pages={513--566},
  year={2023},
  publisher={Springer}
}

@article{masood2023deepfakes,
  title={Deepfakes generation and detection: State-of-the-art, open challenges, countermeasures, and way forward},
  author={Masood, Momina and Nawaz, Mariam and Malik, Khalid Mahmood and Javed, Ali and Irtaza, Aun and Malik, Hafiz},
  journal={Applied intelligence},
  volume={53},
  number={4},
  pages={3974--4026},
  year={2023},
  publisher={Springer}
}

@article{sanchez2015toward,
  title={Toward a universal synthetic speech spoofing detection using phase information},
  author={Sanchez, Jon and Saratxaga, Ibon and Hernaez, Inma and Navas, Eva and Erro, Daniel and Raitio, Tuomo},
  journal={IEEE Transactions on Information Forensics and Security},
  volume={10},
  number={4},
  pages={810--820},
  year={2015},
  publisher={IEEE}
}

@inproceedings{pani2023voice,
  title={Voice Morphing: Two Identities in One Voice},
  author={Pani, Sushanta K and Chowdhury, Anurag and Sandler, Morgan and Ross, Arun},
  booktitle={2023 International Conference of the Biometrics Special Interest Group (BIOSIG)},
  pages={1--6},
  year={2023},
  organization={IEEE}
}

@inproceedings{shen2018natural,
  title={Natural tts synthesis by conditioning wavenet on mel spectrogram predictions},
  author={Shen, Jonathan and Pang, Ruoming and Weiss, Ron J and Schuster, Mike and Jaitly, Navdeep and Yang, Zongheng and Chen, Zhifeng and Zhang, Yu and Wang, Yuxuan and Skerrv-Ryan, Rj and others},
  booktitle={2018 IEEE international conference on acoustics, speech and signal processing (ICASSP)},
  pages={4779--4783},
  year={2018},
  organization={IEEE}
}

@inproceedings{kalchbrenner2018efficient,
  title={Efficient neural audio synthesis},
  author={Kalchbrenner, Nal and Elsen, Erich and Simonyan, Karen and Noury, Seb and Casagrande, Norman and Lockhart, Edward and Stimberg, Florian and Oord, Aaron and Dieleman, Sander and Kavukcuoglu, Koray},
  booktitle={International Conference on Machine Learning},
  pages={2410--2419},
  year={2018},
  organization={PMLR}
}

@article{mandalapu2021multilingual,
  title={Multilingual Audio-Visual Smartphone Dataset and Evaluation},
  author={Mandalapu, Hareesh and Reddy, PN Aravinda and Ramachandra, Raghavendra and Rao, Krothapalli Sreenivasa and Mitra, Pabitra and Prasanna, SR Mahadeva and Busch, Christoph},
  journal={IEEE Access},
  volume={9},
  pages={153240--153257},
  year={2021},
  publisher={IEEE}
}

@article{ramachandra2019smartphone,
  title={Smartphone multi-modal biometric authentication: Database and evaluation},
  author={Ramachandra, Raghavendra and Stokkenes, Martin and Mohammadi, Amir and Venkatesh, Sushma and Raja, Kiran and Wasnik, Pankaj and Poiret, Eric and Marcel, S{\'e}bastien and Busch, Christoph},
  journal={arXiv preprint arXiv:1912.02487},
  year={2019}
}

@inproceedings{snyder2018x,
  title={X-vectors: Robust dnn embeddings for speaker recognition},
  author={Snyder, David and Garcia-Romero, Daniel and Sell, Gregory and Povey, Daniel and Khudanpur, Sanjeev},
  booktitle={2018 IEEE international conference on acoustics, speech and signal processing (ICASSP)},
  pages={5329--5333},
  year={2018},
  organization={IEEE}
}

@article{jung2022pushing,
  title={Pushing the limits of raw waveform speaker recognition},
  author={Jung, Jee-weon and Kim, You Jin and Heo, Hee-Soo and Lee, Bong-Jin and Kwon, Youngki and Chung, Joon Son},
  journal={arXiv preprint arXiv:2203.08488},
  year={2022}
}

@article{jung2020improved,
  title={Improved rawnet with feature map scaling for text-independent speaker verification using raw waveforms},
  author={Jung, Jee-weon and Kim, Seung-bin and Shim, Hye-jin and Kim, Ju-ho and Yu, Ha-Jin},
  journal={arXiv preprint arXiv:2004.00526},
  year={2020}
}

@article{desplanques2020ecapa,
  title={Ecapa-tdnn: Emphasized channel attention, propagation and aggregation in tdnn based speaker verification},
  author={Desplanques, Brecht and Thienpondt, Jenthe and Demuynck, Kris},
  journal={arXiv preprint arXiv:2005.07143},
  year={2020}
}

@article{ulyanov2016instance,
  title={Instance normalization: The missing ingredient for fast stylization},
  author={Ulyanov, D},
  journal={arXiv preprint arXiv:1607.08022},
  year={2016}
}

@inproceedings{pariente2020filterbank,
  title={Filterbank design for end-to-end speech separation},
  author={Pariente, Manuel and Cornell, Samuele and Deleforge, Antoine and Vincent, Emmanuel},
  booktitle={ICASSP 2020-2020 IEEE International Conference on Acoustics, Speech and Signal Processing (ICASSP)},
  pages={6364--6368},
  year={2020},
  organization={IEEE}
}

@inproceedings{ravanelli2018speaker,
  title={Speaker recognition from raw waveform with sincnet},
  author={Ravanelli, Mirco and Bengio, Yoshua},
  booktitle={2018 IEEE spoken language technology workshop (SLT)},
  pages={1021--1028},
  year={2018},
  organization={IEEE}
}

@inproceedings{scherhag2017biometric,
  title={Biometric systems under morphing attacks: Assessment of morphing techniques and vulnerability reporting},
  author={Scherhag, Ulrich and Nautsch, Andreas and Rathgeb, Christian and Gomez-Barrero, Marta and Veldhuis, Raymond NJ and Spreeuwers, Luuk and Schils, Maikel and Maltoni, Davide and Grother, Patrick and Marcel, Sebastien and others},
  booktitle={2017 International Conference of the Biometrics Special Interest Group (BIOSIG)},
  pages={1--7},
  year={2017},
  organization={IEEE}
}

@misc{Verispeak,
  author = {Verispeak},
  title = {VeriSpeak Face and Voice Identification},
  howpublished = {\url{https://www.neurotechnology.com/verispeak.html}},
  year = {2024}, 
  note = {[Online; Feb. 2024]}
}

@inproceedings{venkatesh2020influence,
  title={On the influence of ageing on face morph attacks: Vulnerability and detection},
  author={Venkatesh, Sushma and Raja, Kiran and Ramachandra, Raghavendra and Busch, Christoph},
  booktitle={2020 IEEE International Joint Conference on Biometrics (IJCB)},
  pages={1--10},
  year={2020},
  organization={IEEE}
}

@inproceedings{ferrara2022morphing,
  title={Morphing attack potential},
  author={Ferrara, Matteo and Franco, Annalisa and Maltoni, Davide and Busch, Christoph},
  booktitle={2022 International workshop on biometrics and forensics (IWBF)},
  pages={1--6},
  year={2022},
  organization={IEEE}
}

@article{singh2023deep,
  title={Deep composite face image attacks: Generation, vulnerability and detection},
  author={Singh, Jag Mohan and Ramachandra, Raghavendra},
  journal={IEEE Access},
  volume={11},
  pages={76468--76485},
  year={2023},
  publisher={IEEE}
}

\end{document}